\documentclass{article}
\usepackage[utf8]{inputenc}
\usepackage[T1]{fontenc}
\usepackage{hyperref}
\usepackage{hyperref}

\usepackage{blindtext}
\usepackage{subfig}
\usepackage[toc,page]{appendix}
\usepackage{booktabs}
\usepackage{float}
\usepackage{natbib}
\usepackage{graphicx}
\usepackage{adjustbox}
\usepackage{lipsum}  
\usepackage{soul}
\usepackage{amsmath}

\title{Visualizing CoAtNet Predictions for Aiding Melanoma Detection}

\author{\textbf{Daniel Kvak, MSc.}\footnote{Corresponding author: daniel.kvak@carebot.com}\\
	Carebot s.r.o.\\
	Prague, Czech Republic\\\\
	ORCID: 0000-0001-7808-7773}

\usepackage{natbib}
\usepackage{graphicx}

\begin{document}

\maketitle
\begin{abstract}
    Melanoma is considered to be the most aggressive form of skin cancer. Due to the similar shape of malignant and benign cancerous lesions, doctors spend considerably more time when diagnosing these findings. At present, the evaluation of malignancy is performed primarily by invasive histological examination of the suspicious lesion. Developing an accurate classifier for early and efficient detection can minimize and monitor the harmful effects of skin cancer and increase patient survival rates. This paper proposes a multi-class classification task using the CoAtNet architecture, a hybrid model that combines the depthwise convolution matrix operation of traditional convolutional neural networks with the strengths of Transformer models and self-attention mechanics to achieve better generalization and capacity. The proposed multi-class classifier achieves an overall precision of 0.901, recall 0.895, and AP 0.923, indicating high performance compared to other state-of-the-art networks.
\end{abstract}


\newpage
\section{Introduction}
Artificial intelligence (AI) is emerging to assist healthcare professionals with routine tasks such as removing noise, analysing images or reading medical reports. \citep{hamet2017artificial} In deep learning, currently the most widely adopted AI technique, computer algorithms learn using backpropagation to predict outcomes based on large data sets. \citep{albawi2017understanding} The efficiency of these methods has improved dramatically in recent years and can now be found in areas ranging from computer-aided diagnostics (CADx) to online shopping to autonomous vehicles. However, deep learning tools also raise troubling questions because they solve problems in ways that humans cannot always observe. \citep{wang2020should, holzinger2019causability} There is a growing call among researchers and institutions to clarify the basis on which artificial intelligence makes decisions. \citep{kvak2022towards, amann2020explainability, samek2019towards}

The US Food and Drug Administration (FDA) recently outlined ten guiding principles that should be the cornerstone for the development of clinically applicable artificial intelligence. \citep{us2021good} These guiding principles can help support the introduction of objective, safe and effective medical devices to the market. Beyond monitoring or defining the correct use, the core principles include many practices that have proven successful in other sectors; however, the greatest emphasis is on the so-called explainability of predictions (XAI, explainable artificial intelligence), which limits the risk of clinical bias. \citep{ghassemi2021false}

\section{Background}
One of the most common methods used to identify melanoma is the ABCD rule which was introduced in 1985. \citep{nachbar1994abcd} The acronym stands for Asymmetry, Borderline Irregularity, Changes in Color and Diameter. In 2004, the letter E was added to the ABCD acronym to stand for Evolving. \citep{jensen2015abcdef} Each criterion has certain features that are recognized to distinguish between benign and malignant melanoma. In addition, the method failed to recognize certain malignant nevi in their early stages. \citep{carli2002clinically, liu2005features}

\begin{figure}[!h]
\centering
\includegraphics[width=1\textwidth]{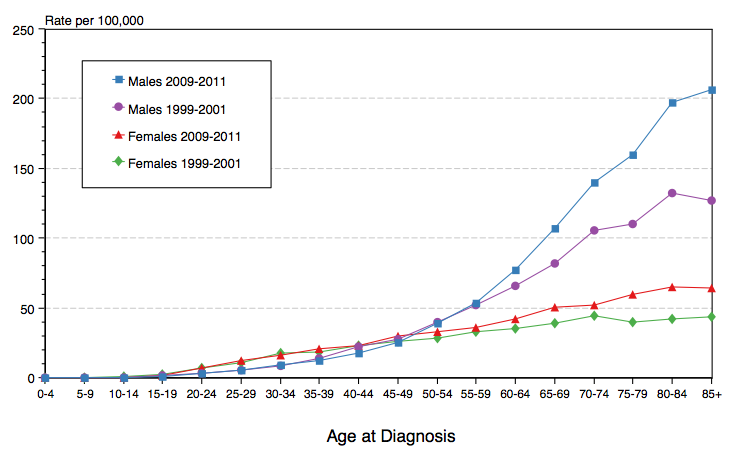}
\caption{\textbf\scriptsize{The delay-adjusted incidence and observed incidence of melanoma by age and gender in the United States between 1975 and 2011.}}
\end{figure}

Melanoma is less common than other types, but it is the most dangerous form of skin cancer because it can spread quickly to other parts of the body. \citep{coit2009melanoma} It results from neoplastic proliferation of melanocytes. Malignant melanoma predominantly affects the skin, but can also affect eyes, ears, leptomeninges, and the mucous membranes of the mouth or genital tract. \citep{bastian2014molecular} The incidence of melanoma is increasing, affecting mainly the light skin population. \citep{matthews2017epidemiology, rigel1996incidence} The pathophysiology of melanoma development is not yet clearly understood. \citep{hida2020elucidation} Multiple pathogenetic mechanisms of melanoma development are hypothesized. Melanoma develops not only on sun-exposed skin, where UV radiation is the main pathogenetic factor, but also in body parts that are relatively protected from radiation. \citep{apalla2017skin, coit2009melanoma} When melanoma is suspected, it is important to biopsy the suspicious lesion on the skin or mucosa (excision with a 1-3 mm margin of tissue) and subsequent histological examination. \citep{bastian2014molecular}

\section{Computational approach}
CADx approaches based on deep learning and computer vision may represent an effective and, above all, affordable alternative to invasive histological examination. \citep{kassani2019comparative} Applications based on convolutional neural networks (CNN) show promising results in medical image detection, classification and segmentation. \citep{li2014medical, anwar2018medical} High accuracy is now achieved in interstitial lung disease classification \citep{shen2015multi} or in the detecion of colorectal adenomas and neoplastic lesions. \citep{yu2016integrating} Many attempts have been made in the literature to improve the performance of CNN, either by using optimization methods to select significant features or by using image preprocessing techniques before the classification step. \citep{thoma2017analysis}

\subsection{Proposed model arcitecture}
CoAtNet offers a unique combination of \textbf{depthwise convolutions} \eqref{eqn1} and \textbf{self-attention} \eqref{eqn2} to allow fast and accurate advancement for large-scale image recognition and classification. The proposed architecture is based on the observation that CNNs tend to exhibit improved generalization (i.e., the difference in performance between training and testing) due to their inductive bias, whereas self-attention models tend to show greater capacity (i.e., the ability to fit large-scale training data). \citep{dai2021coatnet}
\begin{figure}[!h]
\centering
\includegraphics[width=1\textwidth]{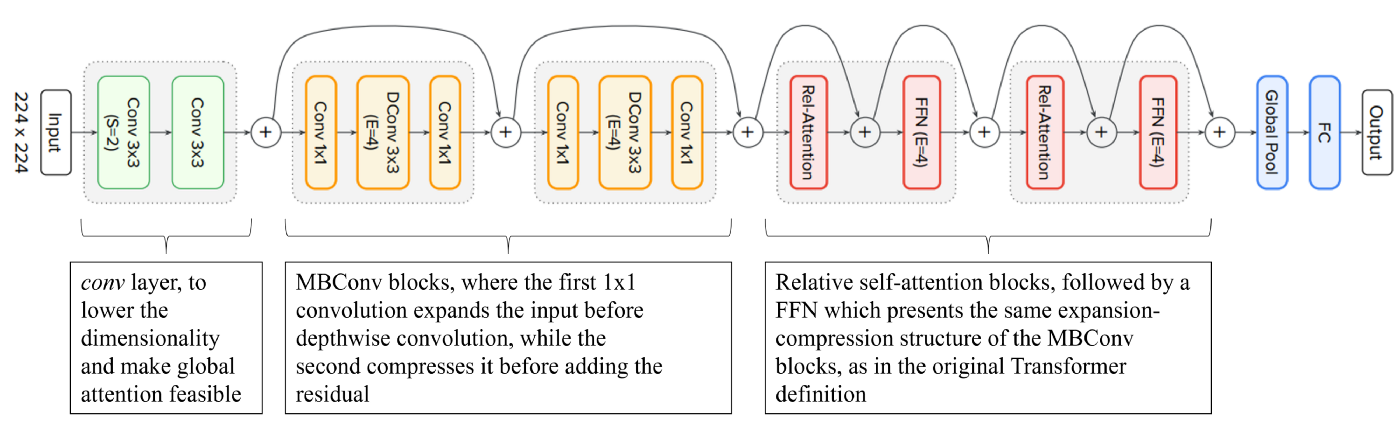}
\caption{\textbf\scriptsize{Overview of the used CoAtNet model. }}
\end{figure}

\eqref{eqn1} is a type of convolution operation where we use one convolution filter for each input channel. \citep{tan2019mixconv} Unlike spatially separable convolutions, depthwise convolutions work with kernels that cannot be split. \citep{guo2019depthwise} In a conventional 2D convolution performed over multiple input channels, the filter is as deep as the input and allows us to arbitrarily mix channels to generate individual features in the output. \citep{chang2016efficient} In contrast, depthwise convolutions maintain each channel separately. We can express this with the formula below:

\begin{equation} \label{eqn1} 
y_{i}=\sum_{j \in \mathcal{L}(i)} w_{i-j} \odot x_{j}
\end{equation}

\eqref{eqn2} has become widespread technique adopted in natural language processing (NLP), with the fully-attentional Transformer model having largely replaced recurrenr neural networks (RNN) and being used in state-of-the-art language understanding models such as GPT, BERT, and XLNet. This technique allows the receptive field to be entire spatial locations, and computes weights based on renormalized pairwise similarity between pairs: if each pixel in the feature map is treated as a random variable and paring covariances are calculated, the value of each predicted pixel can be enhanced or weakened based on its similarity to other pixels in the image. The participating target pixels are the weighted sum of the values of all pixels, where the weights represent the correlation between each pixel and the target pixel. This can be represented by the following formula:

\begin{equation} \label{eqn2} 
y_{i}=\sum_{j \in \mathcal{G}} \underbrace{\frac{\exp \left(x_{i}^{\top} x_{j}\right)}{\sum_{k \in \mathcal{G}} \exp \left(x_{i}^{\top} x_{k}\right)}}_{A_{i, j}} x_{j}
\end{equation}

\section{Dataset}
The development of robust CADx systems for the automated diagnosis of skin lesions is hindered by the small size of clinically evaluated dermatoscopic image datasets available. \citep{garg2021decision} We assembled dermatoscopic images from various publicly available repositories while maintaining a representation of different populations, acquired and stored by different modalities. 

The final dataset consists of 6,826 dermatoscopic images, representative of all important diagnostic categories in the field of various lesions: actinic keratoses, basal cell carcinoma, benign keratosis-like lesions, dermatofibroma, melanoma, nevus, and vascular lesions (angiomas, angiokeratomas, pyogenic granulomas, and hemorrhages). For a fraction of the images ($\sim$50\%), the ground truth was determined by histopathological examination, while in the remaining images the finding was decided by expert consensus or confirmed by in vivo confocal microscopy. A total of 300 images were extracted from the dataset as a test set (100 melanoma, 100 non-melanoma skin cancer, 100 benign skin lesions). The remaining 6,526 dermatology images were split between the training and validation set in an 80/20 ratio.

\begin{figure}[!h]
\centering
\includegraphics[width=1\textwidth]{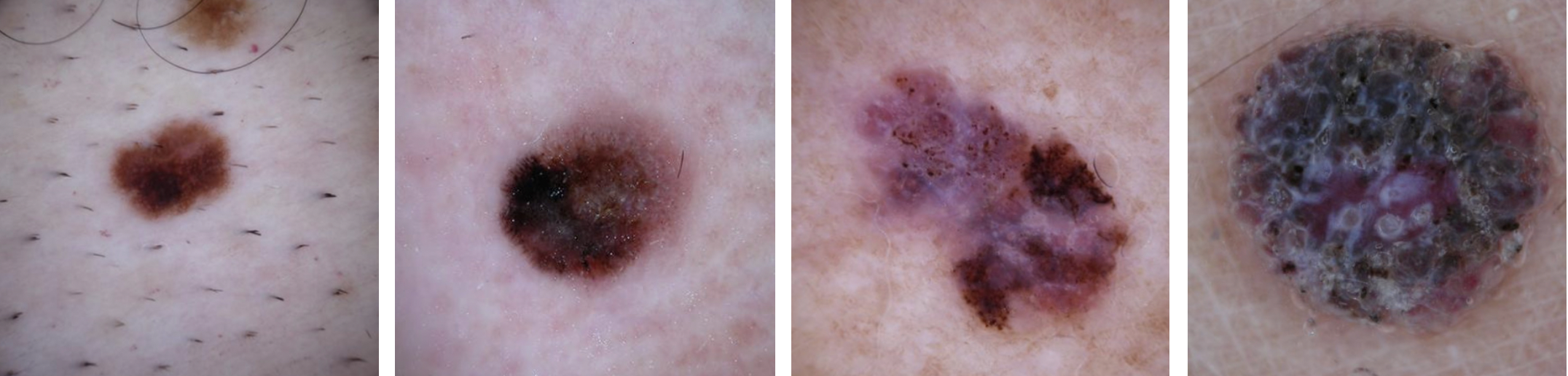}
\caption{\textbf\scriptsize{Examples of melanoma at different stages represented in the training set.}}
\end{figure}
\subsection{Data augmentation}
Data augmentation increases the size of the input training data along with the regularization of the model, thus improving the generalization of the training model. \citep{mikolajczyk2018data} It also helps to create new train examples by randomly applying different transformations to the available dataset to reflect the noise in the real data. \citep{shorten2019survey, elgendi2021effectiveness} In this study, we applied transformations involving random rotations (<= 0.25), modifications in contrast (0.9-1.1) and brightness (0.9-1.1), zoom (<= 0.25), and saturation (0.9-1.1). The extension of validation set was not investigated.

\section{Classifier performance}
The classification performance of the proposed model for multi-class problem was evaluated for each component and the average classification performance of the model was calculated. \autoref{tab:table1} includes the precision \eqref{eqn3} and recall \eqref{eqn4} calculated based on the following equations below:

\begin{equation} \label{eqn3} 	 
Precision = \frac{TP}{TP+FP}
\end{equation}
\begin{equation} \label{eqn4} 
Recall = \frac{TP}{TP+FN}
\end{equation}

For specific experiments and given that there is a class imbalance problem, the most reliable metric is the model average accuracy metric, while given that the accuracy is high, the second most important metric is the recall metric for individual classes. \citep{japkowicz2002class} This is due to the importance of correctly identifying true cases that are malignant. AP (Average Precision) \eqref{eqn5} summarizes a precision-recall curve as the weighted mean of precisions achieved at each threshold \citep{yilmaz2006estimating}, with the increase in recall from the previous threshold used as the weight:

\begin{equation} \label{eqn5}
\text{AP} = \sum_n (R_n - R_{n-1}) P_n
\end{equation}

\begin{table}[] 
\begin{tabular}{@{}lllll@{}}
\toprule
Class                              & No. of images & Precision      & Recall         & AP             \\ \midrule
\textbf{Average model performance} & 6,826 & 0.901 & 0.895 & 0.923 \\
Actinic keratoses         & 332           & 0.786          & 0.821          & 0.772          \\
Basal cell carcinoma               & 514           & 0.880          & 0.922          & 0.919          \\
Benign keratosis-like lesion       & 1,099          & 0.894          & 0.877          & 0.903          \\
Dermatofibroma                     & 115           & 0.875          & 0.913          & 0.944          \\
Melanoma                           & 1,563          & 0.870          & 0.875          & 0.908          \\
Nevus                              & 3,061          & 0.935          & 0.913          & 0.958          \\
Vascular lesions                   & 142           & 1.000          & 0.931          & 0.995          \\ \bottomrule
\end{tabular}
\caption{\label{tab:table1}CoAtNet classifier performance on the used dataset.}
\end{table}

\subsection{Visualizing model predictions}
Despite the classifier showing impressive results on standard metrics, from a clinical perspective, it is important for us to determine whether features relevant to skin lesion detection and analysis were extracted during CoAtNet training using backpropagation. \citep{payer2019integrating} As mentioned in the chapter 1 Introduction, medical devices should not serve as "black boxes" but need to provide additional information about how the model arrived at its predictions. \citep{england2019artificial, amann2020explainability, samek2019towards} Gradient-weighted Class Activation Mapping (Grad-CAM) is a method that uses gradient extraction from the last convolutional layer of a neural network to indicate the pixels that contribute most to the model output and the predicted probability of an image belonging to a predefined class. \citep{selvaraju2017grad} The resulting activation map can be plotted over the original image and can be interpreted as a visual tool to identify the regions that the model predicts whether an image belongs to a particular class. \citep{selvaraju2017grad, panwar2020deep}

\begin{figure}[!h]
\centering
\includegraphics[width=1\textwidth]{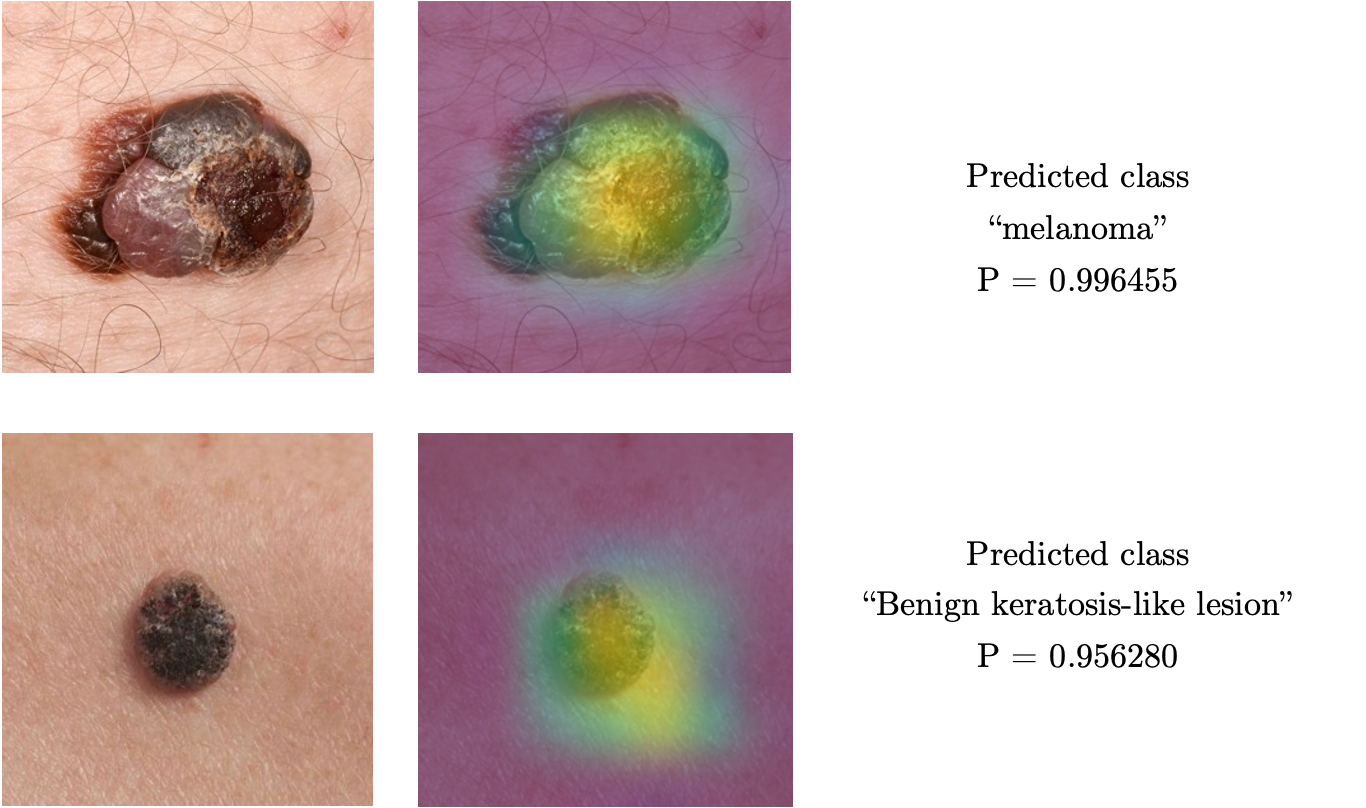}
\caption{\label{fig:fig4}\textbf\scriptsize{Grad-CAM activation heatmap visualization from CoAtNet model on real-world test data.}}
\end{figure}

\subsection{Model performance on test data}
\begin{figure}[!h]
\centering
\includegraphics[width=1\textwidth]{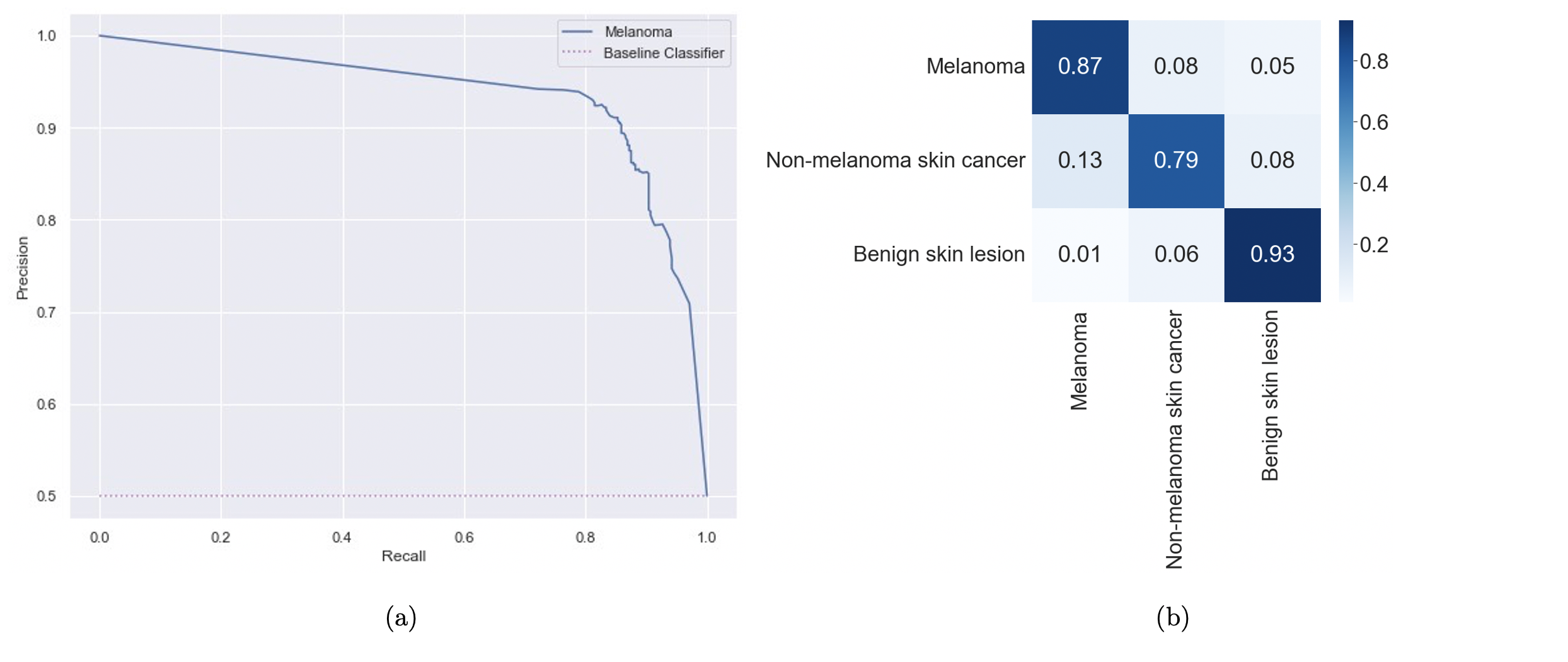}
\centering
\caption{\label{fig:fig5}\textbf\scriptsize{(a) Precision-Recall curve for Melanoma class. (b) Confusion matrix showing the results on the compiled 3-class test set.}}
\end{figure}

The precision-recall curve shows the trade-off between precision and recall for different thresholds. \citep{buckland1994relationship} A high area under the curve represents both high recall and high precision, with high precision associated with low False Positive cases and high recall associated with low False Negative cases. \citep{boyd2013area} 
The combination of the \autoref{fig:fig4} and \autoref{fig:fig5} for the test set suggests that the model learned appropriate features for classification across malignant and benign lesions from a limited dataset.

\section{Conclusions}
In this study, we classified nine skin lesions with a particular focus on melanoma, which, although not as prevalent, is responsible for three-quarters of skin cancer related deaths. The classification of melanoma was performed using no lesion segmentation or complex image preprocessing. The proposed method is based on the state-of-the-art CoAtNet architecture, which incorporates the advantages of depthwise convolution and self-attention mechanism. Considering the necessity of large-scale data for efficient training, we applied data augmentation techniques to the existing dataset. Evidence from the exploratory analysis shows that the proposed approach significantly outperforms state-of-the-art models by achieving model average precision of 0.901, recall 0.895 and AP 0.923. 

\section{Authorship statement}
All persons who meet authorship criteria are listed as authors, and all authors certify that they have participated sufficiently in the work to take public responsibility for the content, including participation in the concept, design, analysis, writing, or revision of the manuscript. Furthermore, each author certifies that this material or similar material has not been and will not be submitted to or published in any other publication.

\section{Ethical procedure}
The authors hereby declare that this research article meets all applicable standards with regards to the ethics of experimentation and research integrity. The authors also declare that the text of the article complies with ethical standards, the anonymity of the patients was respected. 

\newpage
\bibliographystyle{agsm}
\bibliography{references}
\addcontentsline{toc}{chapter}{Bibliography} 

\end{document}